\documentstyle[11pt]{article}
   \newcommand {\nc}{\newcommand}
   \nc{\eq}{\begin{equation}}
   \nc{\en}{\end{equation}}
   \nc{\eqa}{\begin{eqnarray}}
   \nc{\ena}{\end{eqnarray}}
   \nc{\eqann}{\begin{eqnarray*}}
   \nc{\enann}{\end{eqnarray*}}
   \def\prf{{\bf{Proof:}}\\}
   \def\endprf{${\bf{\Box}}$\\}
   \newtheorem{lemma}{Lemma}
   
   \newtheorem{proposition}{Proposition}

   \nc {\dfn}[1]{{\it{#1}}}
   \nc{\nn}{\nonumber}
   \def\dlt{\delta}
   \def\ep{\epsilon}
   
   \def\rep{representation}
   \def\repv{representation }

   \def\etal{{\it et al} }

   \nc{\sqt}{\sqrt{2}}
   \nc{\tsqt}{2\sqt}
   \nc{\msqt}{$\sqt$}
   \nc{\nsqt}{$-\sqt$}
   \nc{\mtsqt}{$\tsqt$}
   \nc{\ntsqt}{$-\tsqt$}

   \def\ot{\otimes}
   
   \def\smdp{>\hspace{-0.2cm}\lhd}

   \nc {\inv}[1]{#1^{-1}}
   \nc {\hc}[1]{{#1}^\dag}
   \nc {\cc}[1]{{#1}^\ast}
   \nc {\ad}[2]{Ad_{#1}({#2})}
   \nc {\wad}{\widetilde{Ad}}
   \nc {\wadf}[2]{\widetilde{Ad}_{#1}({#2})}
   \nc {\pb}[1]{{#1}^\ast}
   \def\tld{\tilde}
   \def\pr{\prime}

   \def\intg{{\cal Z}}
   
   \def\complex{{\cal C}}

   \def\Hs{{\cal H}}
   \def\Jk{{\cal J}}
   \def\St{{\cal S}}
   
   \nc {\ga}[2]{{#1}[{#2}]}
   \nc {\cga}[1]{\ga{\complex}{#1}}
   \nc {\cgag}{\cga{G}}
   \nc {\eu}[1]{E^{#1}}
   \nc {\euf}{\eu{4}}
   \nc {\eun}{\eu{n}}
   \nc {\zn}[1]{\intg^{#1}}
   \nc {\zf}{\zn{4}}
   \nc {\zt} {Z_2}
   \nc {\ztn}[1]{\zt^{#1}}
   \nc {\ztt} {\ztn{2}}
   \nc {\ztth} {\ztn{3}}
   \nc {\ztf} {\ztn{4}}
   \nc {\ztnf}{\ztn{n}}
   \nc {\per}[1]{S_{#1}}
   \nc {\pern}{\per{n}}
   \nc {\irrr}[2]{IRR_{#1}({#2})}
   \nc {\irrc}[1]{\irrr{\complex}{#1}}
   \nc {\irrcg}{\irrc{G}}

   \nc {\mn}{(-)}
   \nc {\mns}[1]{\mn^{#1}}
   \nc {\mo}{(-1)}
   \nc {\mos}[1]{\mo^{#1}}
   \nc {\unit} {{\bf 1}}
   \nc {\unt} {\unit_{2\times 2}}
   \nc {\unth} {\unit_{3\times 3}}
   \nc {\unf} {\unit_{4\times 4}}
   \nc {\une} {\unit_{8\times 8}}
   \nc {\unw} {\unit_{12\times 12}}
   \nc {\zrt} {{\bf 0}_{2\times 2}}
   \nc {\zrth} {{\bf 0}_{3\times 3}}
   \nc {\zrf} {{\bf 0}_{4\times 4}}
    \def\CDalign#1{\bgroup\vcenter\bgroup\tabskip 2pt 
      \baselineskip 14pt \lineskip 3pt \lineskiplimit 3pt
      \halign\bgroup &\hfill$##$\hfill\crcr
      #1\crcr\egroup\egroup\egroup} 
    
  \nc{\act}[3]{S_{{#1}}^{{#2}}[{#3}]}
  \nc{\cact}[2]{S_{Cl}^{{#1}}[{#2}]}
  
   \nc {\prj}[4]{\pi_{#1,#2}#2\ot_{S_#1}e^o_{#3,#4}}
   \nc {\prjf}[2]{\prj{o}{#1}{\eta}{#2}}
   \nc {\prjff}{\prjf{h}{i}}
   \nc {\Prj}[4]{\Pi_{#1,#2;#3,#4}}
   \nc {\Prjf}[2]{\Prj{o}{#1}{\eta}{#2}}
   \nc {\Prjff}{\Prjf{h}{i}}
   \nc {\orbt}[2]{\pi_{#1,#2}}
   \nc {\orbtf}[1]{\orbt{o}{#1}}
   \nc {\orbte}{\orbtf{e}}
   \nc {\rp}[3]{\orbt{#1}{#2}(#3)}
   \nc {\rpf}[2]{\rp{o}{#1}{#2}}
   \nc {\stbrp}[2]{D^o_#1(\tld{s}(#2))}
   \nc {\stbrpme}[4]{\stbrp{#1}{#2}^{#3}_{#4}}
   \nc {\stbrpf}[1]{\stbrp{\eta}{#1}}
   \nc {\stbrpmef}[3]{\stbrpme{\eta}{#1}{#2}{#3}}
   \nc {\ch}[2]{\chi_{#1;#2}}
   \nc {\che}[3]{\ch{#1}{#2}(#3)}
   \nc {\chf}[1]{\ch{o}{#1}}
   \nc {\chff}{\chf{\eta}}
   \nc {\chef}[2]{\che{o}{#1}{#2}}
   \nc {\cheff}[1]{\chef{\eta}{#1}}
   \nc {\chsb}[1]{\chi^o_{#1}}
   \nc {\chsbe}[2]{\chsb{#1}(#2)}
   \nc {\chsbf}{\chsb{\eta}}
   \nc {\chsbef}[1]{\chsbe{\eta}{#1}}
   \nc {\cg}[1]{O_{#1}}
   \nc {\cgn} {\cg{n}}
   \nc {\oh} {\cg{4}}
   \nc {\ohd} {{\overline{\oh}}}
   \nc {\ztfb} {\overline{\ztf}}
   \nc {\iso}[2]{ISO_{#1}(#2)}
   \nc {\isod}[1]{\iso{d}{#1}}
   \nc {\eg}[1]{ISO(#1)}
   \nc {\egn}{\eg{n}}
   \nc {\fs}[1]{F(#1)}
   \nc {\fsf}{\fs{\emb}}
   \nc {\fx}[1]{I(#1)}
   \nc {\fxf}{\fx{\emb}}
   \nc {\wrn}[1]{\ztn{#1}\smdp\per{#1}}
   \nc {\wn}{\wrn{n}}
   \nc {\wnn}{\pern^{\zt}}
   \nc {\w}[1]{\per{#1}^{\zt}}
   \nc {\fix}[1]{\per{(n-#1)}\ot \per{p}}
   \nc {\fixp}{\fix{p}}
   \nc {\cb}[1]{C_{#1}}
   \nc {\cn}{\cb{n}}

   \nc {\prm}{\sigma}
   \nc {\prmt}{\tld{\prm}}
   \nc {\prmp}{\prm^\pr}
   \nc {\cyc}[2]{\tau_{{#1}{#2}}}
   \nc {\emb}{\iota}
   \nc {\epy}{\tld{\ep}}
   \nc {\de}[1]{d_{E^{#1}}}
   \nc {\den}{\de{n}}
   \nc {\dy}{\tld{d}}
   \nc {\repw}[2]{e_{(#1)#2}}
   \nc {\repww}[4]{\repw{#1}{#2}\ot\repw{#3}{#4}}
   \nc {\prw}[6]{\pi_{#1,#2}#2\ot_{F_#1}(\repww{#3}{#4}{#5}{#6})}
   \nc {\dm}[1]{d_{({#1})}}
   \nc {\sls}{{\it slash}}
   \nc {\slsv}{{\it slash} }
   \nc {\lcl} {{\it local}}
   \nc {\lclv} {{\it local} }
   \nc {\drln}[5]
    {\put(#1,#2){\line(#3,#4){#5}}}
   \nc {\ybxa}[4]
    {
     \begin{picture}(40,10)
      \drln {0}{0}{0}{1}{10}
      \drln {0}{0}{1}{0}{40}
      \drln {10}{0}{0}{1}{10}
      \drln {0}{10}{1}{0}{40}
      \drln {20}{0}{0}{1}{10}
      \drln {30}{0}{0}{1}{10}
      \drln {40}{0}{0}{1}{10}
      \drln {0}{0}{1}{1}{#1}
      \drln {10}{0}{1}{1}{#2}
      \drln {20}{0}{1}{1}{#3}
      \drln {30}{0}{1}{1}{#4}
     \end{picture}
    }
   \nc {\ybxb}[4]
    {
     \begin{picture}(30,20)
      \put(0,0){\line(0,1){20}}
      \put(0,0){\line(1,0){10}}
      \put(10,0){\line(0,1){20}}
      \put(0,10){\line(1,0){30}}
      \put(0,20){\line(1,0){30}}
      \put(20,10){\line(0,1){10}}
      \put(30,10){\line(0,1){10}}
      \drln {0}{0}{1}{1}{#1}
      \drln {0}{10}{1}{1}{#2}
      \drln {10}{10}{1}{1}{#3}
      \drln {20}{10}{1}{1}{#4}
     \end{picture}
    }
   \nc {\ybxc}[4]
    {
     \begin{picture}(20,20)
      \put(0,0){\line(0,1){20}}
      \put(0,0){\line(1,0){20}}
      \put(10,0){\line(0,1){20}}
      \put(0,10){\line(1,0){20}}
      \put(0,20){\line(1,0){20}}
      \put(20,0){\line(0,1){20}}
      \drln {0}{0}{1}{1}{#1}
      \drln {10}{0}{1}{1}{#2}
      \drln {0}{10}{1}{1}{#3}
      \drln {10}{10}{1}{1}{#4}
     \end{picture}
    }
   \nc {\ybxd}[4]
    {
     \begin{picture}(20,30)
      \drln {0}{0}{0}{1}{30}
      \drln {0}{0}{1}{0}{10}
      \drln {10}{0}{0}{1}{30}
      \drln {0}{10}{1}{0}{10}
      \drln {0}{20}{1}{0}{20}
      \drln {0}{30}{1}{0}{20}
      \drln {20}{20}{0}{1}{10}
      \drln {0}{0}{1}{1}{#1}
      \drln {0}{10}{1}{1}{#2}
      \drln {0}{20}{1}{1}{#3}
      \drln {10}{20}{1}{1}{#4}
     \end{picture}
    }
   \nc {\ybxe}[4]
    {
     \begin{picture}(10,40)
      \drln {0}{0}{0}{1}{40}
      \drln {10}{0}{0}{1}{40}
      \drln {0}{0}{1}{0}{10}
      \drln {0}{10}{1}{0}{10}
      \drln {0}{20}{1}{0}{10}
      \drln {0}{30}{1}{0}{10}
      \drln {0}{40}{1}{0}{10}
      \drln {0}{0}{1}{1}{#1}
      \drln {0}{10}{1}{1}{#2}
      \drln {0}{20}{1}{1}{#3}
      \drln {0}{30}{1}{1}{#4}
     \end{picture}
    }
   
   \nc{\op}{orientation-preserved}
   \nc{\opv}{orientation-preserved }
   \nc{\on}[1]{SO_{#1}}
   \nc{\ohn}[1]{O_{#1}}
   \nc{\of}{\on{4}}
   \nc{\ohf}{\ohn{4}}
   \nc{\ofd}{\overline{\of}}
   \nc{\ohfd}{\overline{\ohf}}
   \nc{\cir}[1]{e^{i({#1})\pi}}
   
   \def\pthv{Pythagoras' Theorem }
   
   \def\dopv{Dirac operator }

   \nc {\norm}[1]{\parallel{#1}\parallel}
   
   \def\DCv{Dirac-Connes }
   \nc{\groupsum}[1]{{\sum_{#1}}^{\pr}}
   \nc{\reducedsum}[1]{{\sum_{#1}}^{\pr\pr}}
 \title{\DCv Operator on Discrete Abelian Groups and
 Lattices}
 \author{Jian Dai\thanks{daijianium@yeah.net},
  Xing-Chang Song\thanks{songxc@ibm320h.phy.pku.edu.cn}\\
  Theoretical Group, Department of Physics, Peking University\\
  Beijing, P.R.China, 100871}
 \date{January 5th, 2001}
 
\begin{document}
  \begin{titlepage}
  \maketitle
  \begin{abstract}
  \noindent
   A kind of \DCv operator defined in the framework of Connes' NCG is introduced on
   discrete abelian groups; it satisfies a Junk-free condition, and
   bridges the NCG composed by Dimakis, M\"{u}ller-Hoissen and Sitarz
   and the NCG of Connes. Then we apply this operator to
   d-dimensional lattices.
  \end{abstract}
  \end{titlepage}
  \section{Introduction}
   In a series of works, A. Dimakis and F. M\"{u}ller-Hoissen(D\&M) developed the noncommutative
   geometry(NCG) on discrete sets, whose foundation
   is a universal differential calculus
   as well as its {\it reduction} on a discrete set \cite{DMs1}\cite{DMs2}\cite{DM1}\cite{DM2}.
   When a discrete set is endowed with a group structure,
   D\&M's NCG coincides with the NCG devised by A. Sitarz \cite{sitarz1}\cite{sitarz2}.
   In this case, the differential calculus and reduction can be formulated
   using the
   left-invariant forms \cite{FB}. With the merits being simple and intuitive mathematically, this approach of NCG is
   essentially a cohomological description of {\it broken lines} on discrete
   sets; therefore, it provides a natural framework to describe physical systems on
   discrete sets, i.e. classical or quantum fields on lattices \cite{DMa1}. In fact, D\&M
   deduced the correct Wilson action of gauge field on lattices
   within their formalism in \cite{DMa1}.
   \footnote
   {D\&M applied their geometry in the exploration of integrable systems also, on which
   we will not touch in this article \cite{DMa2}.}
   However, neither D\&M nor Sitarz paid much attention to the ``fermionic'' contents on discrete sets which
   would give some insight into the famous puzzle of chiral fermions on lattices
   \cite{nogo} from a NCG point of view, though D\&M have discussed
   the ``\rep s'' of their geometry in \cite{DMs1}\cite{DM1}\cite{DMb1}.\\

   On the other hand, A. Connes finished formulating the axioms
   for his NCG after works
   \cite{co0}\cite{co1}\cite{VB1}\cite{co2}\cite{co3} being
   accumulated.
   This approach of NCG essentially describes the classical
   differential geometry using the tools of operator algebra and
   generalizes this description into the realm of noncommutative
   algebras; despite of the requirement of a tremendous mathematical preparation,
   it has an intimate relation with ``fermionic'' contents of
   nature, for its key concept, a generalized \dopv to which we will refer as
   \DCv opertor in this paper, determines the metric
   structure on a noncommutative space.\\

   In this work, we compose a kind of \DCv operator for
   discrete abelian groups, thus we determine a Connes' geometry as a spinor \repv
   of D\&M-Sitarz' geometry on these groups.
   This operator can be regarded as a generalization of
   the so-called ``naturally-defined'' \dopv for lattice fermions
   in our previous work \cite{ds1}.
   This article is organized as follows. We review briefly
   different versions of NCG in Section \ref{sec1}. Then we explore our
   \DCv operator in Section \ref{sec2}. The ``naturally-defined'' \dopv
   is introduced in Section \ref{sec3}. Some discussions
   are put in Section \ref{sec4}. Being deserved to be mentioned,
   some other authors have also attempted to introduce new
   intuitions into solving the problem of geometric description of spinor on discrete systems from NCG point of
   view. J. Vaz generalized Clifford algebra to be non-diagonal in spacetime \cite{v}. Balachandran
   \etal studied a solution in discrete field theories based on the fuzzy sphere and its
   Cartesian products \cite{fuzzy}.
  \section{Noncommutative Geometries}
  \label{sec1}
   We first introduce the concept of universal graded differential
   algebra $\Omega(A)=\oplus_{k=0}^\infty\Omega^k(A)$ upon a unital associative algebra
   $A$ with $\Omega^0(A)=A$.
   $\Omega(A)$ is a free algebra generated by symbols $\{a,da:a\in A\}$ subject
   to the relations $d\unit=0$, $d(a_0a_1)-da_0a_1-a_0da_1=0$. The
   latter will enable us to express elements in $\Omega(A)$ as
   linear combinations of monomials of the form $a_0da_1...da_n$. A differential
   $d$ is defined by relations $d(a_0da_1...da_n)=da_0da_1...da_n$,
   $d(da_0da_1...da_n)=0$ which are equivalent to the requirement that
   $d^2=0$, and one can check that $d$ satisfies a graded Leibnitz
   law
   \[
     d(\omega_p\omega^\pr)=(d\omega_p)\omega^\pr+(-)^p\omega_p(d\omega^\pr),
     \forall \omega_p\in \Omega^p(A),\forall \omega^\pr\in
     \Omega(A)
    \]
   \subsection{NCG on discrete sets, Reduction of Differential Calculus}
    Now let $A$ be the algebra of all complex functions on a discrete set $S$
    whose elements are labeled by a subset
    of integer and denote by latin characters
    $i,j,k,...$. $A$ has a {\it natural basis} $\{e^i; i\in S,
    e^i(j)=\dlt_j^i\}$ such that any function $f$ in $A$ can be decomposed as
    $f=\sum_{i\in S}f(i)e^i$. The algebraic structure of $A$ can be expressed as
    \[
     e^ie^j=e^i\dlt^{ij}, \forall i,j\in S
    \]
    and the unit is $\unit=\sum_{i\in S}e^i$.
    D\&M proved that under this circumstance, the universal
    differential algebra or differential calculus $(\Omega(A),d)$
    on $S$ satisfies that
    \begin{lemma}\label{lemmaDM}
     1)Let $e^{ij}=e^ide^j, i\neq j$, then $\{e^{ij},i\neq j\}$ is a
     basis of $\Omega^1(A)$;\\
     2)$e^{ij}e^{kl}=e^{ij}e^{jl}\dlt^{jk}$;\\
     3)$e^{i_1...i_r}:=e^{i_1i_2}e^{i_2i_3}...e^{i_{r-1}i_r}, i_k\neq i_{k+1},k=1,2,...,r-1$
     form a basis for $\Omega^{r-1}(A), r=3,4,...$;\\
     4)$e^{i_1...i_p}e^{i_{p+1}...i_r}
     =e^{i_1...i_pi_{p+2}...i_r}\dlt^{i_pi_{p+1}}, p=1,2,...,r-p=1,2,...$;\\
     5)$de^{i_1i_2...i_r}=\sum_{k=0}^{r+1}\sum_{j\neq i_k,j\neq i_{k+1}}
     {(-)^{k}e^{i_1...i_kji_{k+1}...i_r}}$, for $r=1,2,...$;\\
     6)The cohomology groups of $(\Omega(A),d)$ is trivial.
    \end{lemma}
    The geometric interpretation of $e^{i_1...i_r}, r=1,2,...$ is
    the algebraic dual of a
    (r-1)-step broken line $i_1...i_r$; therefore, Lemma \ref{lemmaDM} has a simple and
    natural geometric picture on discrete sets.\\

    A reduction or a reduced differential algebra of $(\Omega(A),d)$
    is defined to be $\Omega(A)/{\cal I}$ in which the
    two-sided ideal ${\cal I}$ is generated by a specific subset of
    $\{e^{ij}\}$; in another word, $\Omega(A)$ is reduced to a
    more meaningful differential calculus modulo relations generated by
    setting $e^{ij}$ in this subset to be zero.
   \subsection{NCG on discrete groups, Reduction of Left-invariant 1-Forms}
    First we introduce some notations.
    We will use $G$ instead of $S$ for a discrete set equipped with a group
    structure, denote its elements by $g, h,...$ and write the unit of $G$ as $e$.
    For all $g\in G$,
    $\bar{g}:=g^{-1}$. Let $G^\pr=G\backslash\{e\}$, and $\groupsum{g}:=\sum_{g\in
    G^\pr}$. The right(left)-translations induced by group multiplications on
    $A$ are defined to be $(R_gf)(h)=f(hg),(L_gf)(h)=f(gh)$.
    Formally we write $\partial_gf=R_gf-f$. As for abelian groups, $R_g=L_g=:T_g$.\\

    Left-invariant 1-forms in $\Omega^1(A)$ are defined as
    \eq\label{def}
     \chi^g=\sum_{h\in G}e^hde^{hg}, \forall g\in G^\pr
    \en
    as well as $\chi^e=-\groupsum{g}{\chi^g}$ for convenience, such that
    \eq\label{diff}
     df=\groupsum{g}\partial_gf\chi^g
    \en
    \begin{lemma}(Sitarz)
     Using left-invariant 1-forms only without appealing to the definition in Eq.(\ref{def}),
     one can show that all the requirements on a differential calculus, linearity,
     nilpotent, graded Leibnitz rule, are guaranteed sufficiently and necessarily,
     if that $\chi^g f=(R_gf)\chi^g$,
     $d\chi^g+\{\chi^e,\chi^g\}+\groupsum{h}_{\neq
     g}{\chi^h\chi^{\bar{h}g}}=0$, together with
     Eq.(\ref{diff}) hold.
    \end{lemma}
    Hence, Eq.(\ref{diff}) can be written as $df=-[\chi^e,f]$.\\

    A left-invariant reduction is generated by specifying a subset of $G^\pr$ denoted as
    $G^{\pr\pr}$ and setting $\chi^g=0, g\in G^\pr\backslash
    G^{\pr\pr}$. Let $\reducedsum{g}:=\sum_{g\in G^{\pr\pr}}$ and
    still $\chi^e=-\reducedsum{g}\chi^g$,
    then Eq.(\ref{diff}) becomes
    $df=\reducedsum{g}\partial_gf\chi^g=-[\chi^e,f]$.
   \subsection{K-Cycles, Junk and Distance Formula}
    Connes defines a {\it K-cycle} to be a triple $(A,\Hs,D)$ consisting of a *-algebra $A$, a
    faithful unitary \repv $\pi$ of $A$ on a Hilbert space $\Hs$ and an (unbounded)
    self-adjoint operator $D$(\DCv operator) on $\Hs$ with compact resolvent, such
    that $[D,\pi(a)]$ is bounded for all $a\in A$. Parallel to
    reductions on discrete sets, an extension of $\pi$ to a
    \repv of the universal differential algebra $\Omega(A)$ on
    $\Hs$ making use of \DCv operator $D$ is required for
    the purpose to introduce a meaningful differential structure
    on $A$. First, extend $\pi$ to be a *-\repv of
    $\Omega(A)$ in $\Hs$ by defining that
    \[
     \pi(a_0da_1...da_n):=\pi(a_0)[D,\pi(a_1)]...[D,\pi(a_n)]
    \]
    Note that, since we do not care the involution property of differential algebra
    in this paper, our definition here omits a "$i^n$" from the
    conventional one for simplicity. Second, to make $\pi$
    be a differential \rep, we define {\it Junk ideal}
    $\Jk^n=ker(\pi|_{\Omega^n(A)})$ and
    \[
     \Omega_D(A)=\oplus_{n=0}^\infty\Omega^n_D(A),
     \Omega^n_D(A):=\pi(\Omega^n(A))/\pi(d\Jk^{n-1})
    \]
    Junk ideal will become nontrivial if $n\geq 2$, namely that one
    has to consider
    $\pi(d\Jk^1)=\{\sum_j[D,\pi(a^j_0)][D,\pi(a^j_1)]:a^j_0,a^j_1\in A,\sum_j\pi(a^j_0)[D,\pi(a^j_1)]=0\}$
    to define $\Omega^2_D(A)$ well.
    \begin{lemma}\label{junkfree}(Sufficient Junk-free condition in second
    order)\\
     If $D^2\in \pi(A)^\pr$ where $\pi(A)^\pr$ is the commutants of
     $\pi(A)$,
     then $\pi(d\Jk^1)=\emptyset$.
    \end{lemma}
    \prf
     The statement can be verified by two identities
     $[a,b][a,c]=\{a,b[a,c]\}-b\{a,[a,c]\}$, $\{a,[a,b]\}=[a^2,b]$.\\
    \endprf
    On the other hand, one can induce a metric $d_D(,)$ on the state space
    $\St(A)$ of $A$ by Connes' distance formula
    \[
     d_D(\phi, \psi)=sup\{|\phi (a)-\psi (a)|: a\in A, \norm{[D,\pi(a)]}\leq
     1\}, \forall \phi, \psi \in \St(A)
    \]
  \section{\DCv Operator on Discrete Abelian Groups}
  \label{sec2}
   In this section, we focus on discrete abelian groups $G$ and try to define
   K-cycles $(A,\Hs,D)$ on them. As a prescription, we suppose that
   the translations $T_g, \forall g\in G$ can be induced from $A$ into $\Hs$, and be denoted
   as $\widehat{T_g}$ which satisfies
   \eq\label{trans}
    \widehat{T_g}\pi(f)=\pi(T_gf)\widehat{T_g}
   \en
   Note that Eq.(\ref{trans}) is obvious if $\Hs$ is a free module
   on $A$. Now we point out that each reduction of $\Omega(A)$
   can be realized by a junk $\Jk^1$; in fact, that $e^{gh}=0$ can
   be implemented by requiring that $\pi(e^{gh})=\pi(e^g)[D,\pi(e^h)]=0$, a constraint on $D$.
   Notice that $\pi(\chi^g)=\sum_{h\in G}{\pi(e^h)[D,\pi(e^{hg})]}=:\eta^g$,
   and implement a left-invariant reduction by setting $\eta^g=0,g\in G^\pr\backslash G^{\pr\pr}$.
   We will just consider finite left-invariant reductions i.e.
   $\sharp(G^{\pr\pr})<\infty$.\\

   One can check that
   \eq\label{eta}
    \eta^g\pi(f)=\pi(T_g f)\eta^g
   \en
   Accordingly, $D$ can be formally written as
   $D=\reducedsum{g}\eta^g$ and there is
   \eq\label{d}
    \pi(df)=[D,\pi(f)]=\reducedsum{g}\pi(\partial_gf)\eta^g
   \en
   Inspired by Eq.(\ref{trans}), we require that $\eta^g$ has the
   factorized form $\eta^g=\Gamma^g \widehat{T_g}, g\in G^{\pr\pr}$(without a summation) in which
   $\Gamma^g\in\pi(A)^\pr$, thus
   \[
    D=\reducedsum{g}{\Gamma^g \widehat{T_g}}
   \]
   and Eqs.(\ref{eta})(\ref{d}) hold.
   Moreover, we require that $D$ satisfies the Junk-free condition
   in Lemma \ref{junkfree}, i.e. $D^2\in \pi(A)^\pr$. To gain a solution, first we need
   partition $G^{\pr\pr}$ into three intersectionless subsets
   $G^{\pr\pr}=\Sigma\coprod T\coprod\bar{T}$ where $\Sigma$ contains all 2-order elements in
   $G^{\pr\pr}$ and $T$, $\bar{T}$ contain other high order elements respecting that if
   $g\in T$ then $\bar{g}\in \bar{T}$; accordingly,
   \[
    D=\sum_{\sigma\in \Sigma}{\Gamma^\sigma T_\sigma}
    +\sum_{t\in T}(\Gamma^t T_t+\Gamma^{\bar{t}} T_{\bar{t}})
   \]
   \begin{proposition}
    If there holds the Clifford
    algebra $Cl(\sharp(G^{\pr\pr}))$
    \eq\label{cl1}
     \{\Gamma^s,\Gamma^t\}=0,\{\Gamma^{\bar{s}},\Gamma^{\bar{t}}\}=0,
     \{\Gamma^s,\Gamma^{\bar{t}}\}=\dlt^{st}
    \en
    \eq\label{cl2}
     \{\Gamma^s,\Gamma^\sigma\}=0,\{\Gamma^{\bar{s}},\Gamma^\sigma\}=0,
     \{\Gamma^\sigma,\Gamma^{\sigma^\pr}\}=2\dlt^{\sigma\sigma^\pr}
    \en
    for all $s,t\in T$, $\bar{s},\bar{t}\in \bar{T}$ and
    $\sigma,\sigma^\pr\in\Sigma$,
    then $D$ is a Junk-free \DCv operator.
   \end{proposition}
   \prf
    A straightforward computation shows that $D^2=(\sharp(\Sigma)+\sharp(T))\unit$ if all
    $\Gamma^g$ subject to the above Clifford relations.\\
   \endprf
   Note that more general solutions can be gained by varying those
   non-vanishing anti-commutation relations in
   Eqs.(\ref{cl1})({\ref{cl2}) by a scalar factor.
  \section{\DCv Operator on Lattices}
  \label{sec3}
   Let $G$ be a d-dimensional lattice as a discrete group $\intg^d$
   where $\intg$ is the integer-addition group, and the elements
   in $\intg^d$ are d-dimensional vectors whose components are
   integers. Define unit vectors to be $\hat{\mu},
   \hat{\mu}(i)=\dlt_{\mu i}, \mu,i=1,2,...,d$ and $(T^\pm_\mu f)(x)=
   f(x\pm\hat{\mu})$. Consider a
   reduction $G^{\pr\pr}=\{\pm\mu,\mu=1,2,...,d\}$, hence $
   df=\sum_{\mu=1}^d
   (\partial^+_\mu f\chi^\mu_++\partial^-_\mu f\chi^\mu_-)$.
   According to the previous Section, \DCv operator on $\intg^d$ is
   \[
    D=\sum_{\mu=1}^d(\Gamma^\mu_+ T^+_\mu +\Gamma^\mu_- T^-_\mu)
   \]
   where $\Gamma^\mu_\pm$ satisfy $Cl(2d)$ relations
   \[
    \{\Gamma^\mu_\pm,\Gamma^\nu_\pm\}=0,
    \{\Gamma^\mu_\pm,\Gamma^\nu_\mp\}=\dlt^{\mu\nu}, \mu,\nu=1,2,...,d
   \]
   In \cite{ds1}, we called this operator(with a modification
   which will be pointed out in the next section)
   ``natural-defined'' \dopv on a lattice and proved in $d=2$ that
   $d_D(,)$ coincides with conventional Euclidean distance on this
   lattice.
  \section{Discussions}
  \label{sec4}
   First, we call the Junk-free condition in second order a geometric square-root
   condition; for being more specific, we write it as
   $D^2=N\unit$ where $N$ is a normalization integer. While $\tld{D}^2=\triangle$ is called a physical square-root
   condition when a laplacian is well-defined. On lattices, they are
   connected by the relation
   \[
    \tld{D}=D-\reducedsum{g}\Gamma^g\unit, \tld{D}=\reducedsum{g}{\Gamma^g\partial_g}
   \]
   Second, M. G\"{o}ckeler and T. Sch\"{u}cker pointed out that in a restricted sense,
   conventional lattice \dopv is not compatible with Connes' geometry, due to the contradiction
   of first-order axiom \cite{latt}. Our work on this aspect is in proceeding.\\

   {\bf Acknowledgements}\\
    This work was supported by Climb-Up (Pan Deng) Project of
    Department of Science and Technology in China, Chinese
    National Science Foundation and Doctoral Programme Foundation
    of Institution of Higher Education in China.
    One of the authors J.D. is grateful to Dr. L-G. Jin in Peking
    University and Dr. H-L. Zhu in Rutgers University
    for their careful reading on this manuscript.
  
 \end{document}